\documentclass{Interspeech2024}
\usepackage{booktabs}
\usepackage{multirow}
\usepackage{adjustbox}
\usepackage{subcaption}
\usepackage{cite}
\usepackage{comment}
\usepackage{amsmath,amssymb,bm}
\usepackage{graphicx}

\usepackage{amsmath}
\usepackage{multirow}
\usepackage[linesnumbered,ruled,vlined]{algorithm2e}
\usepackage{siunitx}
\usepackage{comment}

\newcounter{nodecount}

\usepackage{xspace}
\makeatletter
\DeclareRobustCommand\onedot{\futurelet\@let@token\@onedot}
\def\@onedot{\ifx\@let@token.\else.\null\fi\xspace}

\makeatother

\newcommand{\vect}[1]{{\mbox{\boldmath $#1$}}}

\def\appendixautorefname~#1\null{~#1 \null}

\def\equationautorefname~#1\null{(#1\null)}

\interspeechcameraready

\title{Stream-based Active Learning for Anomalous Sound Detection \\
in Machine Condition Monitoring}

\name[affiliation={1}]{Tuan Vu}{Ho}
\name[affiliation={1}]{Kota}{Dohi}
\name[affiliation={1}]{Yohei}{Kawaguchi}

\address{
  $^1$Hitachi, Ltd., Japan}
\email{tuanvu.ho.zt@hitachi.com, kota.dohi.gr@hitachi.com, yohei.kawaguchi.xk@hitachi.com}


\begin{document}
\maketitle

\begin{abstract}
This paper introduces an active learning (AL) framework for anomalous sound detection (ASD) in machine condition monitoring system. Typically, ASD models are trained solely on normal samples due to the scarcity of anomalous data, leading to decreased accuracy for unseen samples during inference. AL is a promising solution to solve this problem by enabling the model to learn new concepts more effectively with fewer labeled examples, thus reducing manual annotation efforts. However, its effectiveness in ASD remains unexplored. To minimize update costs and time, our proposed method focuses on updating the scoring backend of ASD system without retraining the neural network model. Experimental results on the DCASE 2023 Challenge Task 2 dataset confirm that our AL framework significantly improves ASD performance even with low labeling budgets. Moreover, our proposed sampling strategy outperforms other baselines in terms of the partial area under the receiver operating characteristic score.
\end{abstract}

\section{Introduction}
Anomalous sound detection (ASD) \cite{koizumi2017optimizing, kawaguchi2019anomaly, koizumi2019batch, suefusa2020anomalous, purohit2020deep, wilkinghoff2023using} is a crucial task in various fields such as industrial machinery monitoring and fault detection. By analyzing emitted sound, ASD can effectively identify the condition of target objects, offering a straightforward setup for automated monitoring systems. However, in real-world scenarios, obtaining anomalous sound data during the training phase is often challenging due to the rarity of failure conditions. As a result, ASD is commonly treated as an unsupervised task, where only non-anomalous sound is utilized to train the model. Despite recent advancements leveraging pretrained models on auxiliary tasks such as sound recognition \cite{mezza2023zero}, source separation \cite{shimonishi23_interspeech}, and contrastive learning \cite{hojjati2022self}, unsupervised ASD approaches still struggle to generalize to new and unseen anomalies. This limitation poses a significant challenge, especially in dynamic or evolving environments where existing techniques may exhibit sub-optimal performance. Therefore, there is a critical requirement for ASD systems capable of adapting to changing acoustic environments and detecting previously unseen anomalies without requiring extensive retraining.

Active learning (AL) emerges as a promising solution to address the limitations of traditional anomaly detection methods. Several studies have explored AL principles to enhance the efficiency of sound event detection tasks \cite{shuyang2017active, shuyang2018active, Meire2023}. By selecting informative data points for labeling, AL algorithms facilitate the iterative improvement of detection models while minimizing the need for extensive manual annotation. This adaptive learning approach holds significant potential for enhancing the adaptability and reliability of ASD systems, especially in scenarios with limited labeled data availability. Depending on the approach used for selecting samples for annotation, AL can be divided into pool-based AL \cite{wu2019active, lin2022pool} and stream-based AL\cite{vzliobaite2013active, narr2016stream}. In pool-based AL, samples are drawn from a fixed pool of unlabelled data, typically in batch mode, allowing for a global view of the dataset and efficient processing. On the other hand, stream-based AL operates on continuously arriving data, making real-time decisions on whether to request annotation.

In this paper, we focus on stream-based AL methods for ASD in machine condition monitoring, given its better reaction time compared to pool-based AL methods. We investigate a practical scenario where an initial ASD model, trained on a development dataset, is deployed on a cloud server to identify anomalies within continuous streams of samples. In this scenario, we assume that it is possible to request human annotations for certain samples to further enhance the ASD model performance. However, this aspect has not been addressed in previous research. Therefore, this paper proposes a stream-based AL framework to tackle the challenges of ASD. By emphasizing the benefits of adaptive data selection and model refinement, we demonstrate the potential of AL techniques in enhancing the accuracy and resilience of anomaly detection systems.

The remainder of the paper is structured as follows. Section 2 provides a review of related work in the field of anomalous sound detection. In Section 3, we delve into the details of our proposed AL framework for ASD. Section 4 outlines the experimental setup and presents the results of our evaluation, demonstrating the efficacy of our method compared to baseline approaches. Finally, Section 5 offers concluding remarks and outlines for future research.

\begin{figure}[bt]
    \centering
    \includegraphics[width=0.47\textwidth]{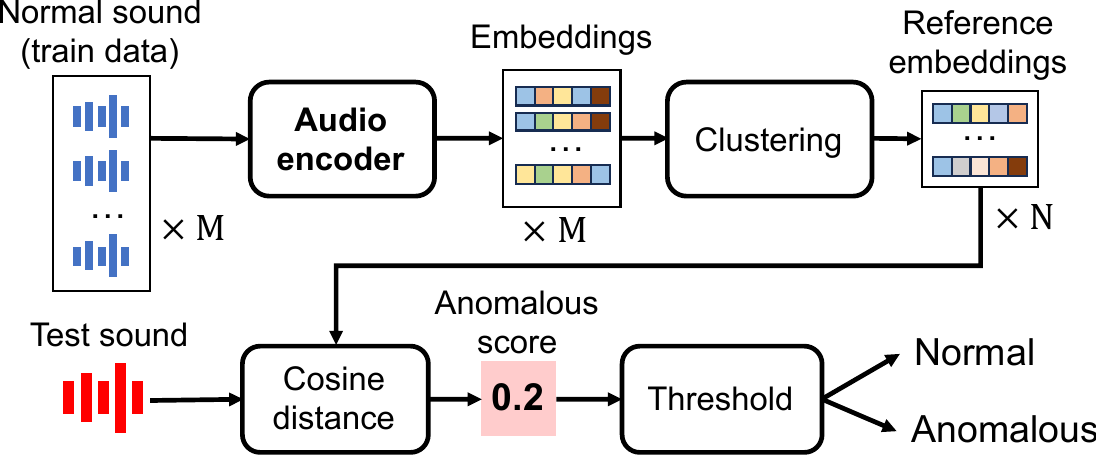}
    \caption{Overview of classification-based ASD system.}
    \label{fig:asd_overview}
\end{figure}

\section{Anomalous Sound Detection Systems}
The ASD task involves identifying and classifying the state of the target object by computing the anomaly score using an anomaly score calculator $A$ with parameters $\theta$. The calculator $A$ is trained to assign higher scores to anomalous samples and lower scores to normal ones. If the anomaly score surpasses a predefined threshold $\phi$, the machine is classified as anomalous, otherwise, it is deemed normal.

The two common approaches for the ASD model are Autoencoder-based and classification-based. The former involves training an Autoencoder model on normal data and determining the anomalous score of test samples by evaluating the reconstruction error \cite{koizumi2017optimizing, kawaguchi2019anomaly, koizumi2019batch, suefusa2020anomalous, purohit2020deep}. Meanwhile, the latter utilizes audio embeddings derived from an audio encoder pretrained on classification task \cite{wilkinghoff2023using, mezza2023zero}. The anomalous score $A(x)$ is estimated by computing the cosine distance between the test embedding and a reference non-anomalous embedding set $\mathcal{N}=\left\{e_j\right\}_{j=1}^J$:

\begin{equation}
\label{eq:anomalous_score}
A(x)= 1 - \max_{e_j \in \mathcal{N}}\frac{\left \langle f(x), e_j \right \rangle}{||f(x)|| ||e_j||},
\end{equation}
where $f(x)$ is the audio encoder, and $\max(\cdot)$ return the max value in the list.  
These reference embedding sets are often constructed through k-means clustering of the non-anomalous
embeddings of training data \cite{wilkinghoff2023using}.

The ASD system utilized in this study is outlined in Figure \ref{fig:asd_overview}. The input features comprise the 513-dimension log-magnitude spectrogram using Short-time Fourier Transform (STFT), where the sampling window is set to 512, the hop size is 128, and the maximum frequency is limited to 8000 Hz. We employ ResNet18 \cite{he2016deep} as the backbone structure for the audio encoder. The audio encoder comprises 4 blocks, each containing 2 stacked residual layers. A temporal statistic pooling layer is utilized to produce the 128-dimensional embedding vector. The network is trained utilizing ArcFace loss \cite{deng2019arcface} with sub-clusters similar to \cite{wilkinghoff2021sub}. 

\begin{figure}[bt]
    \centering
    \includegraphics[width=0.47\textwidth]{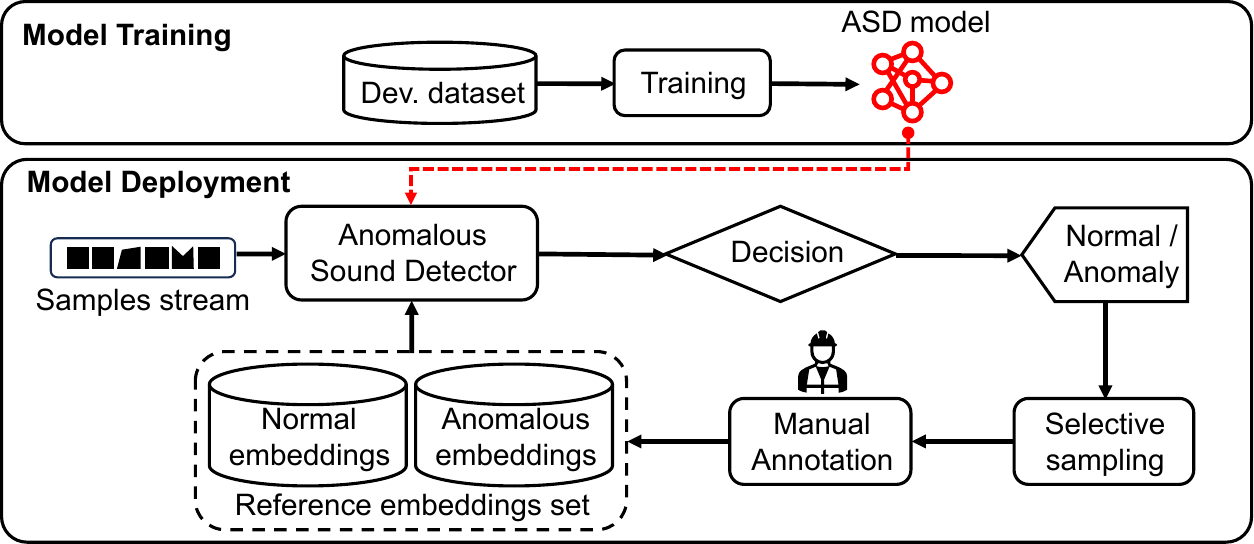}
    \caption{Active learning framework for anomalous sound detection.}
    \label{fig:overview}
\end{figure}

\section{Proposed method}
\subsection{Active learning framework for ASD}
In this section, we introduce our proposed active learning (AL) framework tailored for the ASD task as shown in Figure \ref{fig:overview}.
Given the objective of minimizing costs associated with updating the ASD model, directly modifying the Audio encoder model proves to be challenging due to the high computational expense of retraining neural network models. Moreover, newly annotated data may not be sufficient for effectively training a neural network model.
To avoid this problem, our AL framework focuses on updating the reference embedding set to improve performance while minimizing computation cost. The outline of our active learning framework is shown in Algorithm \ref{alg:online_active_learning}.
Recent studies \cite{koizumi2020spidernet, mezza2023zero} have shown that augmenting the pretrained ASD model with a small number of embeddings from both anomalous and non-anomalous events can enhance performance. Building upon this insight, we adopt the following formulation for the anomalous score:
\begin{gather}
\label{eq:anomalous_score_2}
\hat{A}(x)= \left\{\begin{matrix}
 A(x) & \mathcal{A} = \left\{\varnothing\right\} \\ 
(1-\gamma)A(x) + \gamma S(x) &\mathcal{A}\neq \left\{\varnothing\right\}
\end{matrix}\right.\\
S(x)=\max_{e_i \in \mathcal{A}}\frac{\left \langle f(x), e_i \right \rangle}{||f(x)|| ||e_i||},
\end{gather}
where $A(x)$ is the anomalous score computed by Eq. \ref{eq:anomalous_score}, $S(x)$ is the similarity function between test embedding and the most similar anomalous embedding in the anomalous set $\mathcal{A}=\left\{e_i\right\}_{i=1}^I$.

\begin{algorithm}[t]
    \SetAlgoLined
    \DontPrintSemicolon
    \caption{AL framework for ASD}\label{alg:online_active_learning}
    \label{alg:online_gss}
    \SetAlgoVlined
    \SetNoFillComment
    \SetKw{In}{in}
    \SetKw{Continue}{continue}
    \SetKwInOut{Initialize}{Initialize}
    \KwIn{
    initial reference normal embedding set $\mathcal{N}=\left\{\vect{e}_m\right\}_{m=1}^M$, 
    anomalous embedding set $\mathcal{A}=\left\{\varnothing\right\}$, 
    audio encoder $f\left(\vect{x}\right)$, 
    anomalous score estimator $\hat{A}\left(\vect{e},\mathcal{N}, \mathcal{A}\right)$,
    budget $b\in\left(0,1\right)$, 
    window size to control budget $w>0$, 
    threshold adjustment parameter $\alpha$,
    label query decision $Q\left(s,h,b,w,\alpha\right)$}
    \KwOut{the updated reference non-anomalous and anomalous embedding set.}
    \Initialize{$M \leftarrow$ anomalous score of labeled samples,\\
    $h \leftarrow \mathrm{quantile}(M, 0.9)$\tcp*[f]{90-th percentile of $M$}
    }
    \BlankLine
    \While{sample $\vect{x}$ arrives}{
        $\vect{e}\leftarrow f\left(\vect{x}\right)$\tcp*[f]{Extract embedding}
        $s\leftarrow \hat A\left(\vect{e},\mathcal{N},\mathcal{A}\right)$\tcp*[f]{Estimate anomalous score using Eq. \ref{eq:anomalous_score_2}}\label{algline:uncertainty}
        
        \If{$Q\left(s,h,b,w,\alpha\right)~\text{is}~\mathtt{TRUE}$}{\label{algline:decision}
            Acquire the groundtruth label $y$ of $\vect{x}$\label{algline:update_start}\\
            \If {$y$==1}{
                $\mathcal{A}\leftarrow \mathcal{A} \cup \left\{\vect{e}\right\}$
            }
            \Else{
                $\mathcal{N}\leftarrow \mathcal{N} \cup \left\{\vect{e}\right\}$\\
                Update $M$ 
            }
        }
    }
\end{algorithm}

\subsection{Sampling strategy}
Our approach to sample selection follows the least-certainty-based sampling method \cite{barnabe2015active}. In an unsupervised ASD system, the anomaly score indicates the uncertainty of whether input samples belong to the normal sound class. Additionally, selecting unseen normal instances that differ from the learned class can help explore underrepresented areas. The practical advantage of this strategy becomes apparent in live environments, where false positives can be leveraged to continuously improve the model performance through retraining. However, relying solely on the least-certainty sampling strategy may result in overlooked false negative samples.

To address this issue, we propose a hybrid sampling strategy that combines least-certainty sampling with random sampling. At each step, a random variable $\eta$ is drawn from a uniform distribution $\mathcal{U}(0, 1)$. If $\eta$ exceeds a predetermined parameter $\upsilon $ in the range of $[0, 1)$, the least-certainty strategy is employed to decide the label of s; otherwise, the random strategy is used. To control the labeling budget, we vary the labeling threshold using the similar mechanism of Variable Uncertainty \cite{vzliobaite2013active}. Here we use threshold adjustment parameter $\alpha = 0.01$ in our experiment. The spent labeling budget $\hat b$ is estimated using a moving average window as in \cite{vzliobaite2013active} with window size $w = 200$. The detail of our sampling strategy and budget manager is demonstrated in Algorithm \ref{alg:margin_uncertainty}.

\begin{algorithm}[t]
    \SetAlgoLined
    \DontPrintSemicolon
    \caption{Hybrid certainty-based query decision function $Q\left(s,S,b,w,\alpha\right)$}
    \label{alg:margin_uncertainty}
    \SetAlgoVlined
    \SetKwInOut{Initialize}{Initialize}
    \SetKw{In}{in}
    \SetKw{Continue}{continue}
    \KwIn{Anomalous score of input sample $s$, 
    labeling threshold $h$,
    labeling budget $b$,
    window size $w$,
    threshold adjustment parameter $\alpha$
    }
    \KwOut{Whether to query sample $\lambda \in [\texttt{TRUE}, \texttt{FALSE}]$}
    \Initialize{spent labeling cost $\hat{b}\leftarrow 0$, and store last value of updated parameters.}
    \BlankLine
    \uIf{$\hat{b}<b$}{
        \uIf{$s > h$} {
            $\lambda \leftarrow \texttt{TRUE}$\label{algline:skip2}
        } \Else {\label{algline:skip1}
            $\eta \sim \mathcal{U}(0,1)$\\
            \uIf{$\eta > \hat{b}/b$} {
                $\lambda\leftarrow\mathtt{TRUE}$\label{algline:decrease_threshold}
            } \Else {
                $\lambda\leftarrow\mathtt{FALSE}$\label{algline:increase_threshold}
             }
        }
        $h\leftarrow h \times \left(1+\alpha\right)$\tcp*[l]{Increase labeling threshold}
    \Else{
        $h\leftarrow h \times \left(1-\alpha\right)$\tcp*[l]{Reduce labeling threshold} 
        $\lambda \leftarrow \mathtt{FALSE}$
    }
    }
    $\hat{b}\leftarrow\frac{(w-1)\hat{b}+\lambda}{w}$\tcp*{Update spent cost}\label{algline:approx_cost}
    \Return $\lambda$
\end{algorithm}

\section{Experiments}
\subsection{Dataset}
We utilize the machine sound dataset from DCASE 2023 Challenge Task 2 \cite{dohi2023description}, derived from
the MIMII DG \cite{Dohi2022} and ToyADMOS2 \cite{Harada2021} datasets. The dataset consists of three distinct sets: development, additional set, and evaluation, comprising seven machine types each. These recordings are single-channel audio, ranging from \SI{6}{\second} to \SI{18}{\second} in duration, sampled at 16 kHz, and synthesized from various noise sources. The development set comprises 7,000 training samples and 1,400 test samples for seven machine types, including non-anomalous samples from both the source and target domains. The additional training set introduces seven new machine types with 7,000 training samples. The evaluation set shares the same machine types as the additional training set but lacks condition labels, domain information, and attributes, with each machine having 200 samples.

\subsection{Implementation details}
During the training process, an audio encoder with 4.7M parameters is trained with a batch size of 32 over 10 epochs. We utilize
the implementation of ResNet18 and ArcFace loss from open-source wespeaker library\footnote{\url{https://github.com/wenet-e2e/wespeaker}}. The classification target includes machine type, machine ID, and attributes, resulting in a total of 167 distinct classes from the training data. For each machine type, we select $J=42$ normal samples $e^J_{j=1}$ from the normal training samples. Among these, 10 samples belong to the target domain, while others are the cluster centers of 990 source domain samples obtained through k-means clustering.

As for baseline methods, we utilize random selection and
query-by-committee (QBC) sampling strategies in experiments:
\begin{itemize}
    \item \textbf{Random}: in the random strategy, the labeling probability $u$ is sampled from a uniform distribution $\mathcal{U}(0,1)$ regardless of the input sample. The ground truth label for the sample is queried if $1-u$ is higher than the predefined budget $b$.
    \item \textbf{QBC}: QBC is the common approach for active learning that shows good performance for anomaly detection \cite{pelleg2004active}. QBC uses the diversity of results from $N (>1$) models as an indicator of sample uncertainty:
    \begin{equation}
        u = \frac{1}{N}\sum_{n}^{N} |\hat{s}_n - \bar{s}|,
    \end{equation}
    where $\hat{s}_n$ is the output from the $n$-th model and the $\bar{s}=\frac{1}{N}\sum_n^N \hat{s}_n$. In this experiment, we set $N=10$. Each model comprises a reference normal embedding set derived from $90\%$ of all labeled normal samples. The output of each model is the cosine distance to its respective reference normal embedding set. To control the labeling cost, we use a balancing incremental quantile filter \cite{kottke2015probabilistic}, which can be used in an unbounded uncertainty value case. 
\end{itemize}

\begin{table}[bt]
\centering
\caption{Performance of ASD model without AL. Overall performance is calculated from the harmonic mean across all machines.}
\resizebox{0.47\textwidth}{!}{%
    \begin{tabular}{@{}lcccccc@{}}
    \toprule
    Machine type & \multicolumn{3}{c}{AUC [\%]} & \multicolumn{3}{c}{pAUC [\%]} \\
    \cmidrule(lr){2-4} \cmidrule(l){5-7}
    & Source & Target & Mixed & Source & Target & Mixed \\
    \midrule
    ToyTank & 80.40 & 64.16 & 73.69 & 77.89 & 57.68 & 61.79 \\
    ToyNscale & 74.16 & 94.56 & 82.32 & 56.84 & 79.16 & 68.05 \\
    Vacuum & 98.52 & 99.96 & 98.92 & 92.63 & 99.79 & 95.95 \\
    bandsaw & 79.74 & 79.90 & 70.20 & 58.07 & 57.47 & 55.85 \\
    shaker & 88.68 & 65.01 & 73.59 & 76.29 & 50.40 & 52.40 \\
    ToyDrone & 80.04 & 57.88 & 68.34 & 70.74 & 47.58 & 50.84 \\
    grinder & 68.47 & 60.28 & 70.90 & 67.66 & 47.95 & 61.15 \\
    \midrule
    Harmonic mean & 80.47 & 71.48 & 75.75 & 69.67 & 58.77 & 61.23 \\
    \bottomrule
    \end{tabular}
}
\label{tab:auc_pauc}
\end{table}

\subsection{Evaluation metrics}
Similar to the DCASE 2023 Challenge Task 2, our evaluation uses the area under the receiver operating characteristic curve (AUC) and the partial AUC (pAUC) \cite{mcclish1989analyzing}, focusing on AUC for low false positive rates ($0$ to $0.1$). We evaluate various AL frameworks at different labeling budgets, representing the ratio of queried samples for annotation to the total test samples. We also calculate harmonic means across all machines to summarize each method's performance. For consistency, we use the ground-truth labels and evaluation script from the DCASE 2023 Challenge\footnote{\url{https://github.com/nttcslab/dcase2023_task2_evaluator}}.

\begin{figure*}[bt]
    \centering
    \includegraphics[width=0.98\textwidth]{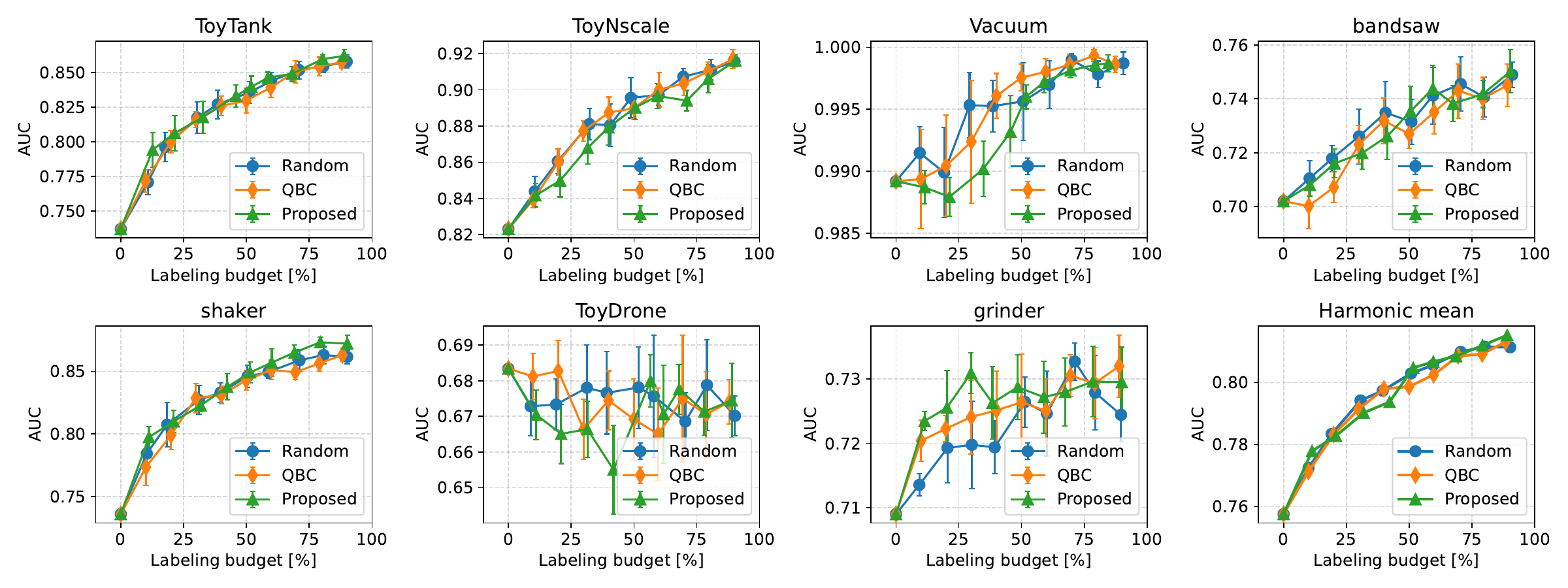}
    \caption{Average AUC scores across 10 runs for each AL method, with error bars representing the $95\%$ confidence interval. The labeling budget denotes the relative ratio of queried samples to the total test samples.}
    \label{fig:auc_random_vs_uncertainty}
\end{figure*}
\begin{figure*}[bt]
    \centering
    \includegraphics[width=0.98\textwidth]{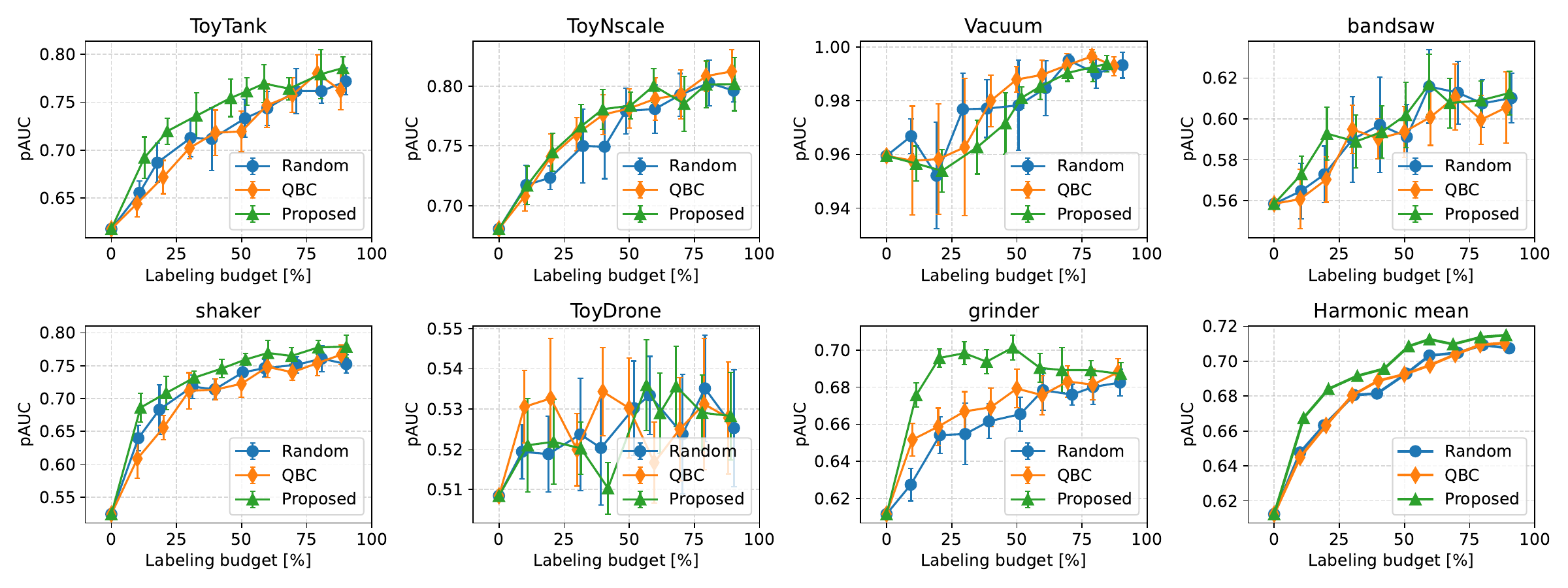}
    \caption{Average pAUC scores across 10 runs for each AL method, with error bars representing the $95\%$ confidence interval. The labeling budget denotes the relative ratio of queried samples to the total test samples.}
    \label{fig:pauc_random_vs_uncertainty}
\end{figure*}
\subsection{Results}
We performed 10 trials for each machine type and labeling budget on the evaluation set. All methods start with the same initial ASD model trained on the development set and additional set.
In each trial, i) we randomly shuffle the order of test samples, and ii) evaluated AUC and pAUC scores in a prequential (interleaved test-then-train) manner for each target machine. Table 1 shows the offline performance of the ASD model without AL. The performance of the initial model is comparable with the top-3 teams in DCASE 2023 Challenge Task 2. Figure 3 illustrates the average AUC score across 10 runs for each machine type using various AL methods at different labeling budgets. Except for ToyDrone, it is evident that all AL methods contribute to an enhancement in the AUC score of ASD as labeling costs increase. Additionally, there is a significant improvement in the average AUC score at a small labeling budget of $10\%$ across all AL methods, indicating the efficacy of AL for ASD. Furthermore, Figure 4 presents the average pAUC score across 10 runs for each machine type using various AL methods at different labeling budgets. While there may not be a clear difference between AL frameworks in terms of AUC, the pAUC score of our proposed method demonstrates superior improvement compared to other baselines, especially for the grinder machine. 
This can be attributed to the sampling strategy of the proposed method, which prioritizes selecting samples that the model is least confident about. By selecting samples where the model certainty is lowest, least-certainty sampling effectively targets unseen non-anomalous samples, hence improving accuracy at a low false positive rate. In overall, experimental results indicate that our proposed method outperforms both the baseline method using QBC and the random labeling strategy in terms of overall performance.

\section{Conclusion}
This study introduces an AL framework tailored for ASD, aiming to streamline ASD model updates for improved performance while minimizing labeling costs. By focusing on updating the reference embedding set instead of directly retraining the neural network model, our AL framework avoids the expensive update cost of retraining the neural network. Through experimental validation, we illustrate the effectiveness of the proposed sampling strategy to enhance ASD performance while maintaining low update costs. As for future work, further exploration of diverse selection criteria and sampling strategies could enhance the efficiency and effectiveness of the AL framework.

\bibliographystyle{IEEEtran}
\bibliography{mybib}

\begin{thebibliography}{10}
\providecommand{\url}[1]{#1}
\csname url@samestyle\endcsname
\providecommand{\newblock}{\relax}
\providecommand{\bibinfo}[2]{#2}
\providecommand{\BIBentrySTDinterwordspacing}{\spaceskip=0pt\relax}
\providecommand{\BIBentryALTinterwordstretchfactor}{4}
\providecommand{\BIBentryALTinterwordspacing}{\spaceskip=\fontdimen2\font plus
\BIBentryALTinterwordstretchfactor\fontdimen3\font minus \fontdimen4\font\relax}
\providecommand{\BIBforeignlanguage}[2]{{%
\expandafter\ifx\csname l@#1\endcsname\relax
\typeout{** WARNING: IEEEtran.bst: No hyphenation pattern has been}%
\typeout{** loaded for the language `#1'. Using the pattern for}%
\typeout{** the default language instead.}%
\else
\language=\csname l@#1\endcsname
\fi
#2}}
\providecommand{\BIBdecl}{\relax}
\BIBdecl

\bibitem{koizumi2017optimizing}
Y.~Koizumi, S.~Saito, H.~Uematsu, and N.~Harada, ``Optimizing acoustic feature extractor for anomalous sound detection based on neyman-pearson lemma,'' in \emph{2017 25th European Signal Processing Conference (EUSIPCO)}.\hskip 1em plus 0.5em minus 0.4em\relax IEEE, 2017, pp. 698--702.

\bibitem{kawaguchi2019anomaly}
Y.~Kawaguchi, R.~Tanabe, T.~Endo, K.~Ichige, and K.~Hamada, ``Anomaly detection based on an ensemble of dereverberation and anomalous sound extraction,'' in \emph{ICASSP 2019-2019 IEEE International Conference on Acoustics, Speech and Signal Processing (ICASSP)}.\hskip 1em plus 0.5em minus 0.4em\relax IEEE, 2019, pp. 865--869.

\bibitem{koizumi2019batch}
Y.~Koizumi, S.~Saito, M.~Yamaguchi, S.~Murata, and N.~Harada, ``Batch uniformization for minimizing maximum anomaly score of dnn-based anomaly detection in sounds,'' in \emph{2019 IEEE Workshop on Applications of Signal Processing to Audio and Acoustics (WASPAA)}.\hskip 1em plus 0.5em minus 0.4em\relax IEEE, 2019, pp. 6--10.

\bibitem{suefusa2020anomalous}
K.~Suefusa, T.~Nishida, H.~Purohit, R.~Tanabe, T.~Endo, and Y.~Kawaguchi, ``Anomalous sound detection based on interpolation deep neural network,'' in \emph{ICASSP 2020-2020 IEEE International Conference on Acoustics, Speech and Signal Processing (ICASSP)}.\hskip 1em plus 0.5em minus 0.4em\relax IEEE, 2020, pp. 271--275.

\bibitem{purohit2020deep}
H.~Purohit, R.~Tanabe, T.~Endo, K.~Suefusa, Y.~Nikaido, and Y.~Kawaguchi, ``Deep autoencoding gmm-based unsupervised anomaly detection in acoustic signals and its hyper-parameter optimization,'' \emph{arXiv preprint arXiv:2009.12042}, 2020.

\bibitem{wilkinghoff2023using}
K.~Wilkinghoff and F.~Fritz, ``On using pre-trained embeddings for detecting anomalous sounds with limited training data,'' in \emph{2023 31st European Signal Processing Conference (EUSIPCO)}.\hskip 1em plus 0.5em minus 0.4em\relax IEEE, 2023, pp. 186--190.

\bibitem{mezza2023zero}
A.~I. Mezza, G.~Zanetti, M.~Cobos, and F.~Antonacci, ``Zero-shot anomalous sound detection in domestic environments using large-scale pretrained audio pattern recognition models,'' in \emph{ICASSP 2023-2023 IEEE International Conference on Acoustics, Speech and Signal Processing (ICASSP)}.\hskip 1em plus 0.5em minus 0.4em\relax IEEE, 2023, pp. 1--5.

\bibitem{shimonishi23_interspeech}
K.~Shimonishi, K.~Dohi, and Y.~Kawaguchi, ``{Anomalous Sound Detection Based on Sound Separation},'' in \emph{Proc. INTERSPEECH 2023}, 2023, pp. 2733--2737.

\bibitem{hojjati2022self}
H.~Hojjati and N.~Armanfard, ``Self-supervised acoustic anomaly detection via contrastive learning,'' in \emph{ICASSP 2022-2022 IEEE International Conference on Acoustics, Speech and Signal Processing (ICASSP)}.\hskip 1em plus 0.5em minus 0.4em\relax IEEE, 2022, pp. 3253--3257.

\bibitem{shuyang2017active}
Z.~Shuyang, T.~Heittola, and T.~Virtanen, ``Active learning for sound event classification by clustering unlabeled data,'' in \emph{2017 IEEE International Conference on Acoustics, Speech and Signal Processing (ICASSP)}.\hskip 1em plus 0.5em minus 0.4em\relax IEEE, 2017, pp. 751--755.

\bibitem{shuyang2018active}
------, ``An active learning method using clustering and committee-based sample selection for sound event classification,'' in \emph{2018 16th International Workshop on Acoustic Signal Enhancement (IWAENC)}.\hskip 1em plus 0.5em minus 0.4em\relax IEEE, 2018, pp. 116--120.

\bibitem{Meire2023}
M.~Meire, J.~Zegers, and P.~Karsmakers, ``Active learning in sound-based bearing fault detection,'' in \emph{Proceedings of the 8th Detection and Classification of Acoustic Scenes and Events 2023 Workshop (DCASE2023)}, Tampere, Finland, September 2023, pp. 111--115.

\bibitem{wu2019active}
D.~Wu, C.-T. Lin, and J.~Huang, ``Active learning for regression using greedy sampling,'' \emph{Information Sciences}, vol. 474, pp. 90--105, 2019.

\bibitem{lin2022pool}
B.~Lin, Z.~Yu, F.~Huang, and L.~Guo, ``Pool-based sequential active learning for regression based on incremental cluster center selection,'' in \emph{2021 Ninth International Conference on Advanced Cloud and Big Data (CBD)}.\hskip 1em plus 0.5em minus 0.4em\relax IEEE, 2022, pp. 176--182.

\bibitem{vzliobaite2013active}
I.~{\v{Z}}liobait{\.e}, A.~Bifet, B.~Pfahringer, and G.~Holmes, ``Active learning with drifting streaming data,'' \emph{IEEE transactions on neural networks and learning systems}, vol.~25, no.~1, pp. 27--39, 2013.

\bibitem{narr2016stream}
A.~Narr, R.~Triebel, and D.~Cremers, ``Stream-based active learning for efficient and adaptive classification of 3d objects,'' in \emph{2016 IEEE International Conference on Robotics and Automation (ICRA)}.\hskip 1em plus 0.5em minus 0.4em\relax IEEE, 2016, pp. 227--233.

\bibitem{he2016deep}
K.~He, X.~Zhang, S.~Ren, and J.~Sun, ``Deep residual learning for image recognition,'' in \emph{Proceedings of the {IEEE} conference on computer vision and pattern recognition}, 2016, pp. 770--778.

\bibitem{deng2019arcface}
J.~Deng, J.~Guo, N.~Xue, and S.~Zafeiriou, ``{ArcFace}: {Additive} angular margin loss for deep face recognition,'' in \emph{Proceedings of the {IEEE/CVF} conference on computer vision and pattern recognition}, 2019, pp. 4690--4699.

\bibitem{wilkinghoff2021sub}
K.~Wilkinghoff, ``Sub-cluster {AdaCos}: {Learning} representations for anomalous sound detection,'' in \emph{2021 International Joint Conference on Neural Networks (IJCNN)}.\hskip 1em plus 0.5em minus 0.4em\relax IEEE, 2021, pp. 1--8.

\bibitem{koizumi2020spidernet}
Y.~Koizumi, M.~Yasuda, S.~Murata, S.~Saito, H.~Uematsu, and N.~Harada, ``{SPIDERnet}: {Attention} network for one-shot anomaly detection in sounds,'' in \emph{ICASSP 2020-2020 IEEE International Conference on Acoustics, Speech and Signal Processing (ICASSP)}.\hskip 1em plus 0.5em minus 0.4em\relax IEEE, 2020, pp. 281--285.

\bibitem{barnabe2015active}
V.~Barnab{\'e}-Lortie, C.~Bellinger, and N.~Japkowicz, ``Active learning for one-class classification,'' in \emph{2015 IEEE 14th international conference on machine learning and applications (ICMLA)}.\hskip 1em plus 0.5em minus 0.4em\relax IEEE, 2015, pp. 390--395.

\bibitem{dohi2023description}
K.~Dohi, K.~Imoto, N.~Harada, D.~Niizumi, Y.~Koizumi, T.~Nishida, H.~Purohit, R.~Tanabe, T.~Endo, and Y.~Kawaguchi, ``Description and discussion on {DCASE} 2023 challenge task 2: {First-shot} unsupervised anomalous sound detection for machine condition monitoring,'' \emph{Proceedings of Detection and Classification of Acoustic Scenes and Events (DCASE) Workshop}, 2023.

\bibitem{Dohi2022}
K.~Dohi, T.~Nishida, H.~Purohit, R.~Tanabe, T.~Endo, M.~Yamamoto, Y.~Nikaido, and Y.~Kawaguchi, ``{MIMII DG}: Sound dataset for malfunctioning industrial machine investigation and inspection for domain generalization task,'' in \emph{Proceedings of the 7th Detection and Classification of Acoustic Scenes and Events 2022 Workshop (DCASE2022)}, Nancy, France, November 2022.

\bibitem{Harada2021}
N.~Harada, D.~Niizumi, D.~Takeuchi, Y.~Ohishi, M.~Yasuda, and S.~Saito, ``{ToyADMOS2}: Another dataset of miniature-machine operating sounds for anomalous sound detection under domain shift conditions,'' in \emph{Proceedings of the Detection and Classification of Acoustic Scenes and Events Workshop ({DCASE})}, Barcelona, Spain, November 2021, pp. 1--5.

\bibitem{pelleg2004active}
D.~Pelleg and A.~Moore, ``Active learning for anomaly and rare-category detection,'' \emph{Advances in neural information processing systems}, vol.~17, 2004.

\bibitem{kottke2015probabilistic}
D.~Kottke, G.~Krempl, and M.~Spiliopoulou, ``Probabilistic active learning in datastreams,'' in \emph{Advances in Intelligent Data Analysis XIV: 14th International Symposium, IDA 2015, Saint Etienne. France, October 22-24, 2015. Proceedings 14}.\hskip 1em plus 0.5em minus 0.4em\relax Springer, 2015, pp. 145--157.

\bibitem{mcclish1989analyzing}
D.~K. McClish, ``Analyzing a portion of the roc curve,'' \emph{Medical decision making}, vol.~9, no.~3, pp. 190--195, 1989.

\end{thebibliography}

\end{document}